# Geant4 – Current and Future :
## A Snowmass 2013 White Paper

Makoto Asai and Richard Mount
*SLAC National Accelerator Laboratory*
*2575 Sand Hill Road, Menlo Park, CA 94025*

The US involvement in Geant4 [1,2] started with its adoption by the BaBar [3] experiment at SLAC in 1997 and the subsequent creation of a group at SLAC supporting BaBar and contributing to Geant4 development. SLAC has provided the leadership of the international Geant4 Collaboration [4] for the recent years and carries major responsibilities for evolutionary development of the existing code with work to implement multithreading and explore new application domains as well as new technologies such as GPUs. This paper presents the current and the future developments being carried by the SLAC Geant4 team.

## 1. Geant4 version 10 – Geant4 with multi-threading capabilities

SLAC leads the collaborative international effort of commissioning Geant4 version 10, a new series of Geant4 with multi-thread capabilities. With the anticipation of the first production release in December 2013, the Geant4 collaboration has released the first beta-release in June 2013 [5].

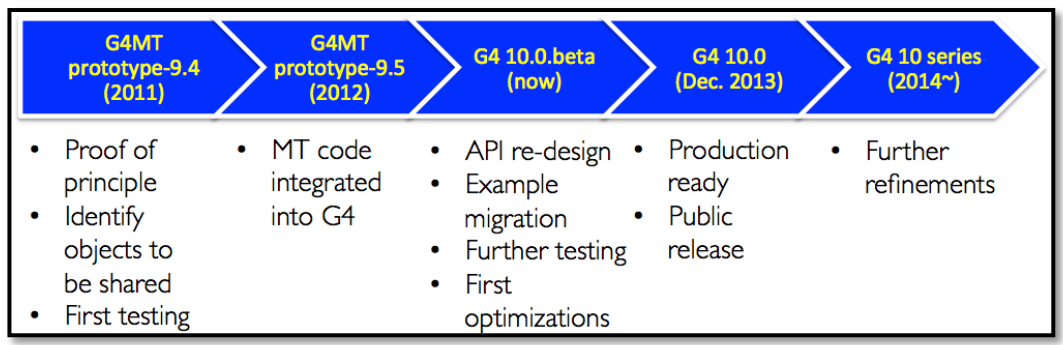

Fig.1. Time evolution of the development of multi-thread in Geant4

Geant4 version 10 offers so-called event-level parallelism, each simulating event runs on separate thread while majority of the memory-consuming part of the code (e.g. geometry, physics tables) are shared over threads. This design choice minimizes the changes of the major Geant4 API, and thus minimizes the migration cost of user applications. Preliminary test of recent pre-beta version with x86_64 and Intel Xeon Phi co-processor has demonstrated a good scalability (Fig.2). Geant4 version 10 uses POSIX standards, allowing easy integration with user-preferred parallelization frameworks such as TBB (Thread Building Block [6]). Version 10 also works with MPI.

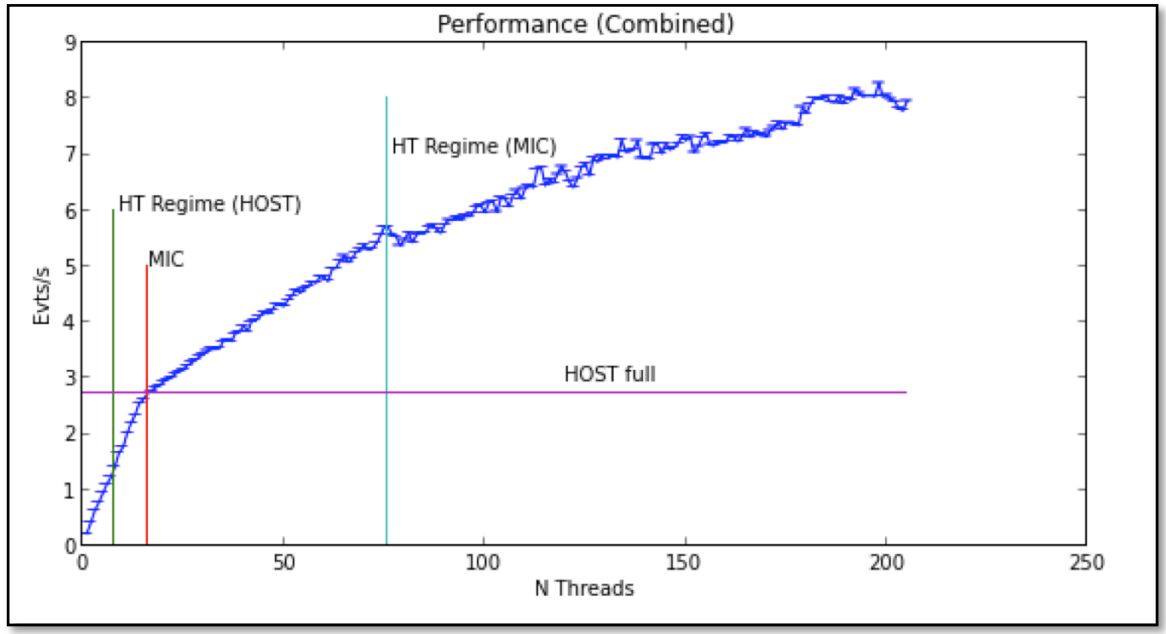

Fig.2. Preliminary scalability test of Geant4 version 10-pre-beta with 8-core Intel Xeon CPU and 1 Intel Xeon Phi co-processor (60 core). From left to right, CPU, CPU in hyper-thread, co-processor, and co-processor in hyper-thread.

## 2. Extending Geant4 capabilities to the new application domains

SLAC is investing in new application domains of Geant4. They include phonon transport in cryogenic crystals [7], electron/hole transport in semiconductors [8], the channeling effect in atomic lattices [9], and the thermal motion of atoms in a vacuum cryostat [10]. These investments address the needs of experiments in which SLAC is participating, and also attract users in various science domains. Fig.3 shows the trajectories of holes in a cryogenic Germanium crystal with phonon emission for the CDMS experiment [11].

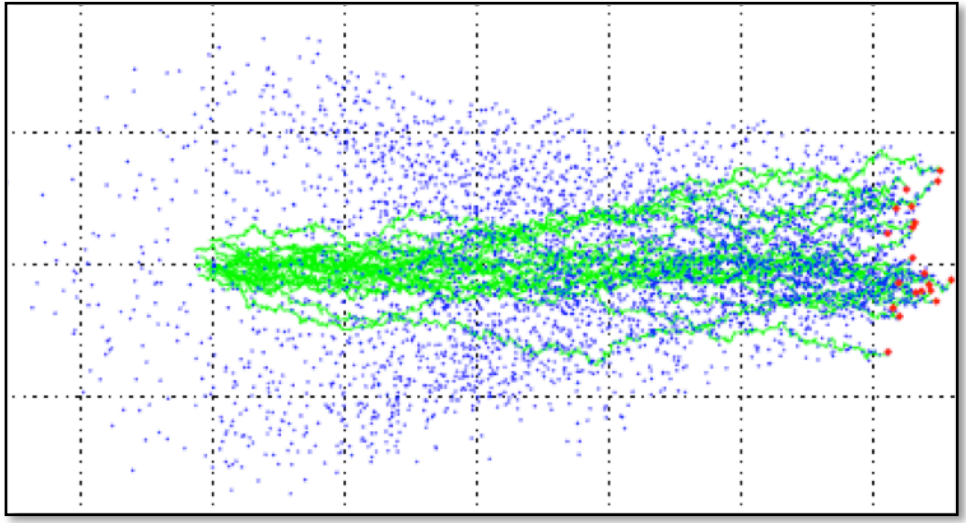

Fig.3. Hole transport in cryogenic Germanium crystal. Red dots represent holes and green lines are their trajectories. Blue dots represent Luke-Neganov phonons.

## 3. Investments in GPU

SLAC is collaborating with ICME (Institute for Computational Mathematics and Engineering) at Stanford and KEK with support from nVidia to study the feasibility of GPUs for particle simulation. The current prototype implements a full set of electromagnetic physics of electron/positron/gamma up to ~100 MeV in regular voxels of water. Its benchmark on TESLA K20 shows $O(10^2)$ speedup compared to the simulation on a single core of i-7 [12].

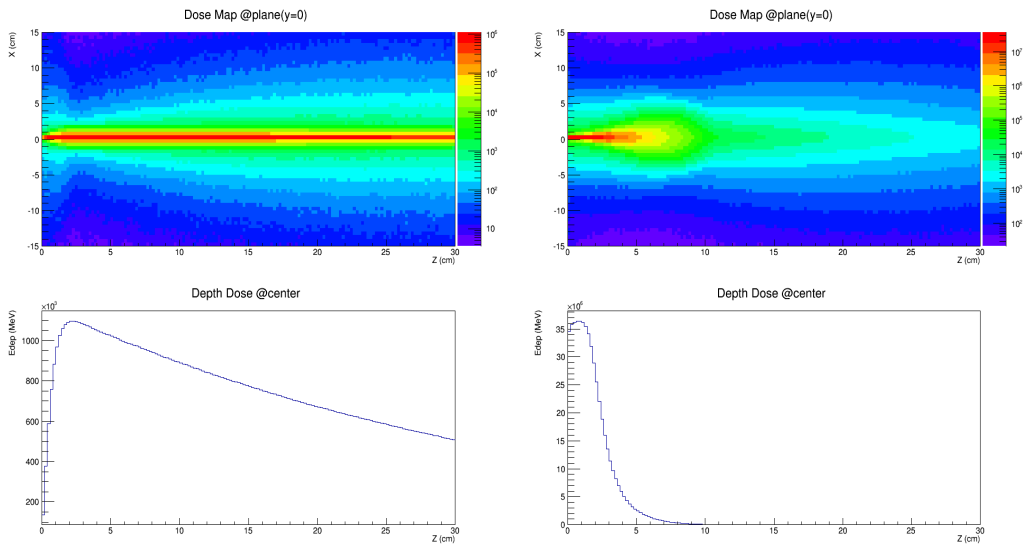

Fig.4. Dose distribution of 6 MeV gamma (left) and 20 MeV electron (right) beams in water simulated on TESLA K20 GPU.